\documentclass[pre,10pt]{revtex4}

\usepackage{amsmath}    % need for subequations
\usepackage{graphicx}   % need for figures
\usepackage{verbatim}   % useful for program listings
\usepackage{color}      % use if color is used in text
\usepackage{subfigure}  % use for side-by-side figures
\usepackage{hyperref}   % use for hypertext links, including those to external documents and URLs

\begin{document}

\title{Fitting the dielectric response of collisionless plasmas by  
       continued fractions}
\author{August Wierling}
\email{august.wierling@uni-rostock.de}
\affiliation{Universit\"at Rostock, Institut f\"ur Physik,
             18051 Rostock, Germany}
\date{\today}

\begin{abstract}
We present an approximation scheme for the dielectric response
of thermal collisionless plasmas at arbitrary degeneracy. A T-fraction
representation is obtained from the known expansions of the real part
of the dielectric function for small and large arguments.
The partial numerators and denominators of the continued fraction
are generated by a modified Q-D algorithm.
For several typical values of the degeneracy parameter $\theta$, extensive
tables for the expansion coefficients and the partial numerators
and denominations are given allowing for an easy implementation of the
fitting function. 
Also, an error analysis is performed.
\end{abstract}

\maketitle

\newcommand{\holter}[2]{\begin{array}{c}\multicolumn{1}{|c}{#1}\\\hline
 \multicolumn{1}{c|}{#2}\end{array}}
\newcommand{\polter}[2]{\begin{array}{c}\multicolumn{1}{c|}{#1}\\\hline
 \multicolumn{1}{|c}{#2}\end{array}}

\section{Introduction}
\label{sec:Introduction}
The dielectric response of a collisionless plasma is an ubiquitous quantity
in plasma physics and in the many-body theory of Coulomb
systems in general. For a fermionic system, it was first 
derived by Lindhard \cite{Lindhard54} and is closely related to the 
dielectric function $\epsilon(k,\omega)$ 
in random phase approximation \cite{Mahan93}. In this way,
the dielectric response is connected to collective effects such 
as screening, plasmons, and Landau damping.
Its knowledge is also important for non-ideal plasmas, since the
effects of non-ideality are typically parameterized by dynamic local
field corrections 
with respect to the ideal response \cite{Mahan93,Ichimaru94}.

A detailed study of the ideal dielectric response for fermionic 
systems has been reported
by Arista and Brandt \cite{Arista84}. Here, we retain this
restriction to fermions, 
although a treatment for bosonic systems is possible in 
a similar manner. In particular, we consider an one-component system
with temperature $T$ and density $n$ interacting by the Coulomb potential. 
Introducing the function $g(x)$ as 
\begin{eqnarray}
  \label{eq:gx}
  g(x) & = & 
  \int_0^{\infty} \frac{y\, dy}{\mbox{exp}\left( y^2/\theta -\eta \right)
      \,+\,1} \, \mbox{ln} \left| \frac{x+y}{x-y} 
  \right|,
\end{eqnarray}
the real part of the dielectric function for fermions is given by
\begin{eqnarray}
  \label{eq:re_eps}
  \mbox{Re}\,\epsilon(k,\omega) & = & 
  1\,+\, \frac{\chi_0^2}{4 z^3} \left[
 g(u+z) -g(u-z) \right] \,\,\,,
\end{eqnarray}
with $u=\omega/(k v_F)$, $z=k/(2 k_F)$, $\chi_0^2=( \pi k_F
a_{\rm B})^{-1}$, and the degeneracy parameter $\theta=k_B T/E_f$.
Here, $E_F,v_F,k_F$ are the Fermi energy $E_F=\hbar^2 k_F^2/(2 m)$, 
Fermi velocity $v_F=\hbar k_F/m$, and 
Fermi wave number $k_F=( 3 \pi^2 n)^{1/3}$, respectively. 
The Bohr radius is denoed by $a_B$,
$\eta$ the chemical potential obtained from $2/3 = F_{1/2}(\eta) 
\,\theta^{3/2}$ with the Fermi function $F_{1/2}$.
The imaginary part of the 
dielectric function is known analytically for all degeneracies 
\begin{eqnarray}
  \label{eq:im_eps}
  \mbox{Im}\, \epsilon(k,\omega) & = & 
  \frac{\pi \chi_0^2}{8 z^3} \theta \,\mbox{ln}
  \left( \frac{ 1\,+\,\mbox{exp}\left[ \eta-\left(u-z\right)^2/\theta \right]
   }{1\,+\,\mbox{exp}\left[ \eta-\left(u+z\right)^2/\theta \right]
   }\right)\,\,\,. 
\end{eqnarray}
For special cases, see the tables in Ref.~\cite{Arista84}.

The dielectric response enters a number of important physical
observables such as optical properties \cite{Griem97}, the single-particle
self-energy in $GW$ approximation \cite{Mahan93}, and the dynamical collision
frequency \cite{Reinholz00}. A fast and reliable computation is therefore
of crucial importance to obtain these quantities. However, a direct 
evaluation of the integral Eq.~(\ref{eq:gx}) is often too slow 
in many applications. Thus, approximative analytical expressions
for $g(x)$ are desirable.

Continued fractions appear in a variety of applications in theoretical
physics \cite{cont_phys}. 
Also, they are of importance in approximation theory and are
closely related to Pad{\'e} approximants which in turn enjoy a 
number of interesting applications in various fields \cite{Baker96,Baker70}.
In particular, there is a close correspondence between T-fractions 
\cite{Thron48} and 
two point Pad{\'e} approximations, see Refs.~\cite{McCabe76}
. Since an expansion of $g(x)$ at 
$z=0$ and at $z=\infty$ is known, the T-fractions representation
suggests itself as a powerful approximation. Also, due to the compact
form of the continued fraction, only a few fitting parameters have to be given.

\section{Approximation scheme}
\label{sec:approximation_scheme}

Following Arista and Brandt \cite{Arista84}, we can consider the
expansion of $g(x)$ for small and large values of $x$ and obtain
\begin{eqnarray}
  \label{eq:gx_small}
  g(x) & = & 2 H_1(\theta) x\, + \, \frac{2}{3} H_2(\theta) x^3\,+\,
             \frac{2}{5} H_3(\theta) x^5 \,+\,\dots\,\frac{2}{2 i-1} H_i(\theta)
             x^{2 i-1} \,+\dots \,\,\, ,\\ 
  g(x) & = & \frac{2}{3 x}\,+\, \frac{\theta^{5/2} F_{3/2}(\eta)}{3
    x^3}\,+\, \frac{\theta^{7/2} F_{5/2}(\eta)}{5 x^5} \,+\,\dots 
  \,+\, \frac{\theta^{l/2+1} F_{l/2}(\eta)}{l x^l}\,+\,\dots 
  \label{eq:gx_large}
\end{eqnarray}
where $F_{\nu}(x)$ is the Fermi integral of order $\nu$, i.e.
\begin{eqnarray}
  \label{eq:Fermi_integral}
  F_{\nu}(x) & = &
  \int_0^{\infty}\!dt\, \frac{t^{\nu}}{1\,+\,{\rm e}^{t-x}}  \,\,\,
\end{eqnarray}
and $i$ being an integer, $l$ an odd number.
Note, that $H_i$ are hyper-singular integrals which have to be
regularized. Details can be found in appendix
\ref{sec:appendix_small_x}. 

Furthermore, the function $g(x)$ is analytically known in the 
non-degenerate limit $\theta \gg 1$ 
\begin{eqnarray}
  \label{eq:gx_non_degenerate}
  g(x) & = & \frac{2}{3} \theta^{-1/2} {\cal D} \left(x/\theta^{1/2}
 \right)\,\,\,,
\end{eqnarray}
and in the highly degenerate limit $ \theta \ll 1$
\begin{eqnarray}
  \label{eq:gx_degenerate}
  g(x) & = & x+ \frac{1}{2} \left( 1-x^2 \right) \,
  \mbox{ln} \left | \frac{1+x}{1-x} \right|\,\,\,.
\end{eqnarray}
${\cal D}(x)$ is the so-called plasma dispersion function 
\cite{Fried61}
\begin{eqnarray}
  \label{eq:dawson}
  {\cal D}(x) & = & \frac{1}{\sqrt{\pi}} \, P\int_{-\infty}^{\infty}
  \!dy\,\frac{{\rm e}^{-y^2}}{x-y} \,\,\,,  
\end{eqnarray}
and 
closely related to the complex error function \cite{Abramowitz72}.
$P \int$ indicates a Cauchy principal value integration.

It has been shown by McCabe and Murphy \cite{McCabe76}, 
that there exists a close connection between a two-point Pad{\'e}
approximant and a T-fraction \cite{Thron48} given in general by
\begin{eqnarray}
  \label{eq:T-fraction}
  & & c_0\,+\,\polter{F_1 z}{1+G_1 z} 
         \,+\,\polter{F_2 z}{1+G_2 z}
         \,+\,\polter{F_3 z}{1+G_3 z} \,+\, \dots\,\,\,, 
\end{eqnarray}
with $F_i,G_i$ being some constants and $z$ being a complex variable.
Here, we follow the notation introduced in Ref.~\cite{Henrici91}.
In particular, once the expansions for $z=0$ and $z=\infty$ are known
to be of the form 
\begin{eqnarray}
  \label{eq:expansion_at_zero}
  \frac{\mu_0}{z} \,+\, \frac{\mu_1}{z^2} \,+\, \frac{\mu_2}{z^3}  
\,+\,\dots  \,+\, \frac{\mu_k}{z^{k+1}} \,\dots\,\,\,,
\end{eqnarray}
and
\begin{eqnarray}
  \label{eq:expansion_at_infinity}
  -\mu_{-1} \,-\,\mu_{-2} z \,-\,\mu_{-3} z^2 \,-\,\dots \,-\,
   \mu_{-k} z^{k-1} \,+\,\dots\,\,\,,
\end{eqnarray}
with expansion coefficient $\mu_i$ and a complex variable $z$,
the T-fraction representation can be generated by a Q-D algorithm
provided that all expansion coefficients are different from zero.
A modified Q-D algorithm has been developed by de Andrade et al. 
\cite{Andrade03} to allow for zero expansion coefficients.

Now, the above given expressions for $g(x)$ are almost of the 
form necessary for applying the T-fraction scheme. We have to take 
care of the infinitesimal small imaginary contribution to 
the expansion at $\infty$
\begin{eqnarray}
  \label{eq:imaginary_g}
  \mbox{Im}\,g(x) & = & 
  \frac{\pi}{2} \,\theta\,\mbox{ln} \left( 1\,+\,
  \mbox{exp} \left( \eta -x^2/\theta \right) \right)\,\,\,.  
\end{eqnarray}
by defining $\tilde g(x) =g(x) -i \mbox{Im} g(x)$ and calculating
the continued fraction representation for $\tilde g(x)$.
After generating the representation, we can go back to the original $g(x)$.

\section{Fitting coefficients}
\label{sec:coefficients}

\begin{figure}[t]
\includegraphics[width=0.8\textwidth]{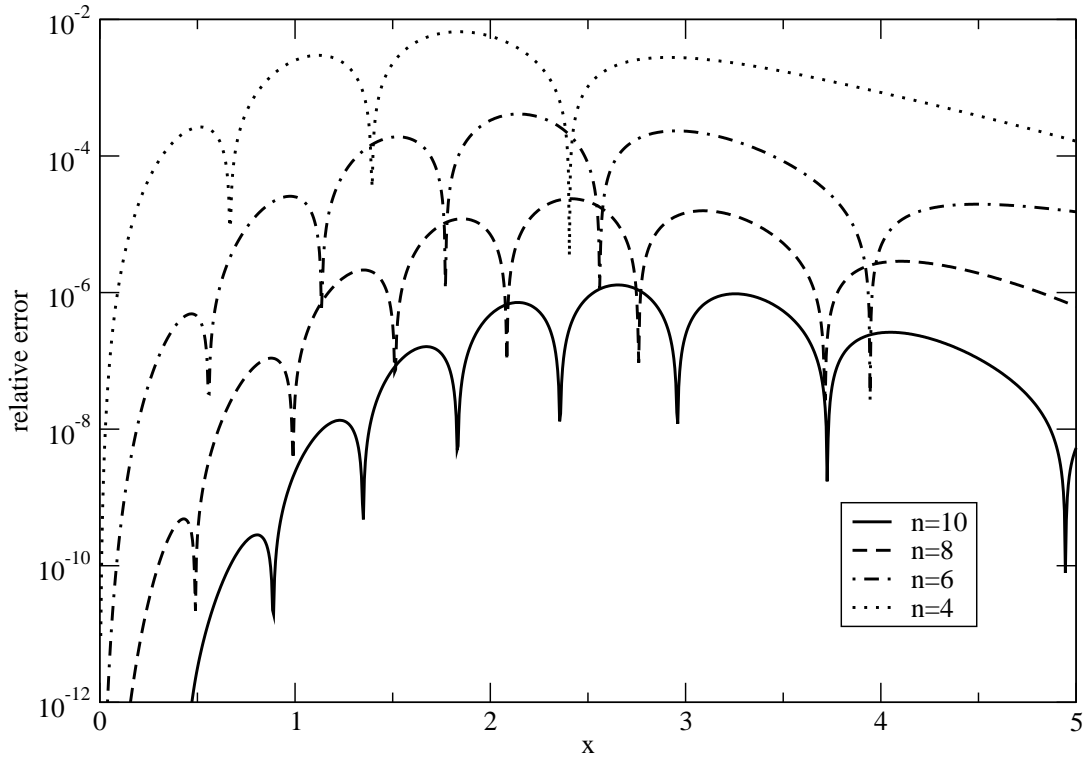}
\caption{The relative error for a T-fraction approximation of the
  function $f_{\infty}(x)=\frac{\sqrt{\pi}}{2} {\rm e}^{-x^2} \left( i \,+\,
  {\rm erfi}(x) \right)$ as a function of the variable $x$ for the
$n$th approximant.}
\label{fig:dawson_error}
\end{figure}

As an illustrative example we study the T-fraction representation
of Dawson's integral, which is closely related to the
plasma dispersion function of a  non-degenerate plasma, cf. 
Eq.~(\ref{eq:dawson}) and see also Ref.~\cite{Fried61}. 
For a detailed discussion of Dawson's integral and its continued
fraction representation see Refs.~\cite{McCabe74}.
The application of T-fractions to this problem has already been considered
by J. McCabe, see Ref.~\cite{McCabe84}. We repeat a few of these
results for illustration.
Since Dawson's integral does not have the proper expansion at
$z=\infty$, we consider the modified function
\begin{eqnarray}
  \label{eq:modified_Dawson}
  f_{\infty}(x) & = & \frac{i \sqrt{\pi}}{2} {\rm e}^{-x^2}\,+\,
  {\cal D}(x)/2 \,=\, \frac{\sqrt{\pi}}{2} {\rm e}^{-x^2} \left( i \,+\,
  {\rm erfi}(x) \right) \,\,\, 
\end{eqnarray}
for a real variable $x$. This function does have the expansions 
\begin{eqnarray}
  \label{eq:modified_dawson_small_x}
  f_{\infty}(x) & = & 
  \frac{i \sqrt{\pi}}{2} + x - \frac{i \sqrt{\pi}}{2} x^2-
  \frac{2}{3} x^3+\frac{i \sqrt{\pi}}{4} x^4+\frac{4}{15}
  x^5-\frac{i \sqrt{\pi}}{12} x^6 -\frac{8}{105} x^7 +
  \frac{i \sqrt{\pi}}{48} x^8 +\frac{16}{945} x^9 -
  \dots\,\,\,, \\ 
  \label{eq:modified_dawson_large_x}
  f_{\infty}(x) & = & \frac{1}{2x }+\frac{1}{4\, x^3}+\frac{3}{8
    \,x^5}+\frac{15}{16\, x^7} +\frac{105}{32\, x^9}+
  \frac{945}{64\, x^{11}}+\dots\,\,\,.
\end{eqnarray}
Using these expansions, one can calculate the T-fraction representation for
$f_{\infty}(x)$ with the help of the modified Q-D algorithm, see 
App.~\ref{sec:appendix_qd_algorithm}.  
In general, excellent convergence is
found. The relative error as a function of the variable $x$ for the
$n$th approximant is shown in
Fig.~\ref{fig:dawson_error}. For $n=10$, the error is better than
$10^{-6}$ for all $x$.

\begin{figure}[t]
\includegraphics[width=0.8\textwidth]{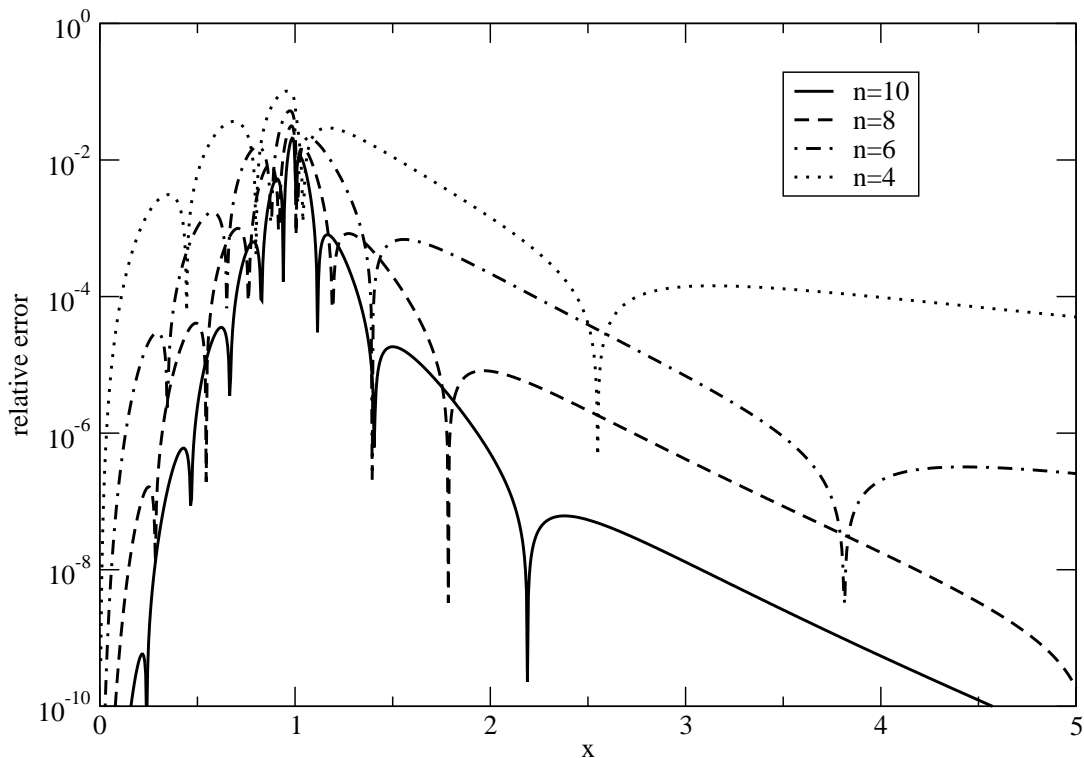}
\caption{The relative error for a T-fraction approximation of the
  function $f_{0}(x)$, see Eq.~(\ref{eq:f_0}), as a function of 
  the variable $x$ for the $n$th approximant.}
\label{fig:entart_error}
\end{figure}

As a second example, we discuss the highly degenerate plasma with
$\theta \to 0$. Again, $g(x)$ as given by Eq.~(\ref{eq:gx_degenerate})
has to be supplemented by an imaginary part to ensure the correct
asymptotic behavior
\begin{eqnarray}
  \label{eq:f_0}
  f_0(x) & = & x+ \frac{1}{2} \left( 1-x^2 \right)\,\mbox{ln}
  \left| \frac{1+x}{1-x} \right|\,+\,i \,\mbox{Im} f_0(x) \,\,\,, 
\end{eqnarray}
with 
\begin{eqnarray}
  \label{eq:f_0_imaginary}
  \mbox{Im} f_0(x) & = & \frac{\pi}{2} \,\times\,\left\{ 
             \begin{array}{cc}
               \left( 1- x^2 \right) & |x| <1 \,\,, \\
               0                     & |x| \ge 1\,\,.
             \end{array}
             \right.
\end{eqnarray}
For this function, the expansions read for $x \to \infty$ and 
$x \to 0$ 
\begin{eqnarray}
  \label{eq:f_0_expansions}
  f_0(x) & \approx &   
   \frac{2}{3\, x} \,+\,\frac{2}{15\, x^3} \,+\,\frac{2}{35\, x^5}
   \,+\,\frac{2}{63\,x^7} \,+\,\dots
\,\,\,,\\
  f_0(x) & \approx & 
  \frac{ i\pi}{2} \,+\,2 x\,-\,\frac{i \pi}{2} x^2\,-\,
  \frac{2}{3} x^3\,-\,\frac{2}{15} x^5\,-\, \frac{2}{35} x^7\,-\,
  \dots
\,\,\,,\nonumber \\  
\end{eqnarray}
respectively.
Again, performing the modified Q-D algorithm leads to partial
numerators and denominators as listed in Tab.~\ref{tab:partial_num_denom}.
As before, the overall convergence is good, which is verified by
inspection of the relative error between the continued fraction
representation and the analytic expression Eq.~(\ref{eq:f_0})
, see Fig.~\ref{fig:entart_error}. Only in the vicinity
of $x=1$, the agreement drops to about $2\%$ for the 10th approximant of
the continued fraction.

\begin{table}[t]
\begin{center}
\begin{ruledtabular}
\begin{tabular}{cccccc}
$\theta$ & $H_1$     & $H_2$ & $H_3$ & $H_4$ & $H_5$ \\ \hline 
  0.1    & 9.914(-1) & -1.0186 & -0.3649 & -0.2482 & -0.2108 \\
  0.2    & 9.611(-1) & -1.0892 & -0.4569 & -0.2997 & -0.1499 \\
  0.3    & 9.081(-1) & -1.1533 & -0.4130 & -0.0950 & 0.1268  \\ 
  0.5    & 7.790(-1) & -1.1037 & -0.1000 & 0.1734  & 0.1217  \\
  0.8    & 6.116(-1) & -0.8336 & 0.1250  & 0.0939  & -0.0075 \\
  1.0    & 5.289(-1) & -0.6692 & 0.1424  & 0.0403  & -0.0133 \\
  1.5    & 3.888(-1) & -0.3970 & 0.0973  & -0.0005 & -0.0038 \\
  2.0    & 3.048(-1) & -0.2552 & 0.0581  & -0.0038 & -0.0007 \\
\end{tabular}
\end{ruledtabular}
\caption{\label{tab:coefficients_H} Expansion coefficients for the 
small $x$ expansion of the function $g(x)$, cf. Eq.~(\ref{eq:gx_small}).}
\end{center}
\vspace{-0.6cm}
\end{table}

\begin{table}[t]
\begin{center}
\begin{ruledtabular}
\begin{tabular}{ccccc}
$\theta$ & $c_{-3}$  & $c_{-5}$   & $c_{-7}$  & $c_{-9}$ \\ \hline 
  0.1    & 1.387(-1) & 6.368(-2) & 3.897(-2) & 2.810(-2) \\
  0.2    & 1.537(-1) & 8.275(-2) & 6.256(-2) & 5.856(-2)  \\
  0.3    & 1.749(-1) & 1.130(-1) & 1.071(-1) & 1.305(0) \\ 
  0.5    & 2.270(-1) & 2.837(-1) & 5.277(-1) & 1.228(0) \\
  0.8    & 3.153(-1) & 4.418(-1) & 8.723(-1) & 2.509(0) \\
  1.0    & 3.771(-1) & 6.042(-1) & 1.565(0) & 5.580(0)  \\
  1.5    & 5.359(-1) & 1.250(0)  & 4.777(0) & 2.532(1) \\
  2.0    & 6.979(-1) & 2.143(0)  & 1.845(1) & 7.639(1) \\
\end{tabular}
\end{ruledtabular}
\caption{\label{tab:coefficients_F} Expansion coefficients for the 
large $x$ expansion of the function $g(x)$, cf. Eq.~(\ref{eq:gx_large}).}
\end{center}
\vspace{-0.6cm}
\end{table}

\begin{table}[t]
\begin{center}
\begin{ruledtabular}
\begin{tabular}{ccccccccccc}
$\theta$ & 0     & 0.1   & 0.2 & 0.3 &  0.5 &  0.8 &  1.0 &  1.5 &
2.0 \\ \hline 
  $b_1$  & 4.244(-1)  & 4.280(-1) & 4.393(-1) & 4.571(-1) & 5.023(-1)
         & 5.749(-1)  & 6.217(-1) & 7.307(-1) & 8.283(-1) \\
  $a_2$  & 4.244(-1)  & 4.280(-1) & 4.393(-1) & 4.571(-1) & 5.023(-1) 
         & 5.749(-1)  & 6.217(-1) & 7.307(-1) & 8.283(-1) \\
  $b_2$  & 9.234(-1)  & 9.407(-1) & 9.900(-1) & 1.061(0)  & 1.224(0)  
         & 1.460(0)   & 1.608(0)  & 1.938(0)  & 2.223(0)  \\
  $a_3$  & 4.522(-1)  & 4.545(-1) & 4.652(-1) & 4.872(-1) & 5.458(-1) 
         & 6.375(-1)  & 6.981(-1) & 8.374(-1) & 9.592(-1) \\
  $b_3$  & 9.613(-1)  & 9.932(-1) & 1.071(0)  & 1.184(0)  & 1.430(0)  
         & 1.731(0)   & 1.953(0)  & 2.413(0)  & 2.782(0)  \\
  $a_4$  & 4.702(-1)  & 4.730(-1) & 4.792(-1) & 5.074(-1) & 5.879(-1) 
         & 6.695(-1)  & 7.673(-1) & 9.678(-1) & 1.117(0)  \\
  $b_4$  & 9.790(-1)  & 1.046(0)  & 1.135(0)  & 1.286(0)  & 1.650(0)  
         & 1.634(0)   & 2.168(0)  & 2.956(0)  & 3.388(0)  \\
  $a_5$  & 4.804(-1)  & 5.072(-1) & 4.800(-1) & 5.116(-1) & 6.525(-1) 
         & 2.700(-1)  & 6.921(-1) & 1.239(0)  & 1.409(0)  \\
  $b_5$  & 9.878(-1)  & 1.223(0)  & 1.177(0)  & 1.260(0)  & 2.046(0)  
         & 3.122(-1)  & 1.545(0)  & 3.720(0)  & 5.740(0)  \\
  $a_6$  & 4.865(-1)  & 7.067(-1) & 4.471(-1) & 3.718(-1) & 9.483(-1) 
         & -4.755(0)  & 7.390(-1) & 2.021(0)  & 3.815(0)  \\
  $b_6$  & 9.923(-1)  & 3.713(0)  & 1.108(0)  & 5.128(-1) & 5.835(0)  
         & -4.275(0)  & -8.132(-1)& 8.054(-1) & -2.160(0) \\
  $a_7$  & 4.902(-1)  & 3.404(0)  & -2.049(-1)& -8.853(-1)& 5.056(1)  
         & 8.397(-1)  & 4.991(0)  & 1.492(-2) & -2.648(0) \\
  $b_7$  & 9.949(-1)  & -4.507(-1)& 4.077(-1) & -6.900(-1)& -8.928(-1)
         & 9.416(-1)  & 5.566(0)  & 7.319(-3) & 2.244(0)  \\
  $a_8$  & 4.926(-1)  & 4.519(-1) & 2.686(0)  & 7.086(-1) & -8.812(-1)
         & -2.490(0)  & -1.334(-3)& 7.783(1)  & 1.698(0)  \\
  $b_8$  & 9.965(-1)  & 1.041(2)  & -2.871(0) & 7.651(-1) & 7.850(0)  
         & -2.190(0)  & -1.653(-3)& 7.817(1)  & -1.833(0) \\
\end{tabular}
\end{ruledtabular}
\caption{\label{tab:partial_num_denom} 
Partial numerators $a_i$ and denominators $b_i$
for the T-fraction approximation of $g(x)$ as a function of the 
degeneracy parameter $\theta$, see Eq.~(\ref{eq:T-fraction_form}).
Note, that $b_1=a_2$.}
\end{center}
\vspace{-0.6cm}
\end{table}

Next, we turn to the general case of arbitrary degeneracy.
Typical expansion coefficients $H_i$ are compiled in 
Tab.~\ref{tab:coefficients_H}. Also, the expansion coefficients
for large $x$ are given essentially by the Fermi integrals for order $l/2$.
They are listed in Tab.~\ref{tab:coefficients_F} for various values
of the degeneracy parameter $\theta$. Finally,
the imaginary contribution due to the modified expression in
Eq.~(\ref{eq:imaginary_g}) has to be determined. It reads
\begin{eqnarray}
  \label{eq:imaginary_g_expansion}
  \mbox{Im}\, g(x) & \approx &
  \mbox{ln} \left( 1+ {\rm e}^{\eta} \right) \,-\, 
  \frac{{\rm e}^{\eta}}{1+{\rm e}^{\eta}} \frac{x^2}{\theta} \,+\,
  \frac{{\rm e}^{\eta}}{2 \left( 1+{\rm e}^{\eta} \right)^2}
  \frac{x^4}{\theta^2} \,+\,
  \frac{{\rm e}^{\eta} \left( {\rm e}^{\eta}-1 \right)}{
  6 \left( 1 + {\rm e}^{\eta} \right)^3} \frac{x^6}{ \theta^3} 
   \nonumber \\ & & 
  +\,\frac{{\rm e}^{\eta} \left( 1- 4 {\rm e}^{\eta}+
                                {\rm e}^{2 \eta}\right)}{
      24 \left( 1 + {\rm e}^{\eta} \right)^4}\frac{x^8}{\theta^4}\,+\,
     \frac{{\rm e}^{\eta} \left( -1 +11 {\rm e}^{\eta} -11 {\rm e}^{2
           \eta}+{\rm e}^{3 \eta} \right)}{
     120 \left( 1 + {\rm e}^{\eta} \right)^5} \frac{x^{10}}{\theta^5}
   \,+\dots \,\,\,.
\end{eqnarray}

Using these coefficients,
the T-fraction approximation is determined using the modified Q-D
algorithm described in App.~\ref{sec:appendix_qd_algorithm}. The
resulting partial numerators and denominators are recorded in 
Tab.~\ref{tab:partial_num_denom}. The general form of the T-fraction
considered here is
\begin{eqnarray}
  \label{eq:T-fraction_form}
  g(x) & = & \mbox{Re} \left[ 
   \polter{\mu_0}{x-i b_1} \,+\, 
     \polter{i a_2 x}{x-i b_2}\,+\, 
     \polter{i a_3 x}{x-i b_3} \,+\, 
     \polter{i a_4 x}{x-ib_4}\,+\, \dots \right]\,\,\,. 
\end{eqnarray}
with $i$ being the imaginary unit, $\mu_0=2/3$.

\section{Conclusions}
\label{sec:conclusions}

It has been shown that 
T-fractions are a powerful method to obtain reliable and 
compact approximative expressions for the dielectric response 
of ideal plasmas at arbitrary degeneracy. In particular, 
a set of eight partial numerators and denominators are sufficient
to obtain a satisfying accuracy for most practical applications.
Fast implementations for the dynamic collision
frequency and the self-energy in $GW$ approximations are possible
based on these results due to a notable acceleration compared to a 
direct evaluation of the integral representation. An extension to 
bosonic systems appears to be straightforward and is work in progress.

\begin{acknowledgments}
This work was supported by the Deutsche Forschungsgemeinschaft within
SFB 652 'Strong correlations and collective effects in radiation
fields'. The author would like to thank Carsten Fortmann and Mathias
Winkel for testing the numerical pay-off  of the final fit formula. 
Also, he would like to thank John McCabe for pointing out 
Ref.~\cite{McCabe84}.
\end{acknowledgments}

\appendix
\section{Small $x$ expansion coefficients}
\label{sec:appendix_small_x}
 
In order to regularize the hyper-singular integrals in Eq.~(\ref{eq:gx_small}),
we consider an expansion of the Fermi function around zero
\begin{eqnarray}
  \label{eq:Fermi_function_zero}
  \frac{1}{\mbox{exp}( y^2/\theta-\eta) +1 } & = & 
  \frac{1}{1+{\rm e}^{-\eta}}\,-\,
  \frac{{\rm e}^{\eta}}{ \left( 1+{\rm e}^{\eta} \right)^2}
  \frac{y^2}{\theta} \,-\,
  \frac{{\rm e}^{\eta} \,\left( {\rm e}^{\eta}-1 \right)}{
  2 \left( 1+{\rm e}^{\eta} \right)^3} \frac{y^4}{\theta^2}\,-\,
  \frac{{\rm e}^{\eta} \,\left( 1-4 {\rm e}^{\eta} +{\rm e}^{2 \eta} \right)}{
  6 \left( 1+{\rm e}^{\eta} \right)^4 }
   \frac{y^6}{\theta^3} \nonumber \\ & & -\,
  \frac{{\rm e}^{\eta} \left( -1 + 11 \eta
       -11 {\rm e}^{2 \eta}+{\rm e}^{3 \eta} \right) 
  \,}{ 24 \left( 1+{\rm e}^{\eta} \right)^5
  }\frac{y^8}{\theta^4}\,+\,\dots \,\,\,.
\end{eqnarray}
The calculation has been done using the computer algebra software
MATHEMATICA \cite{Mathematica6}. In detail, we obtain for the integrands
\begin{eqnarray}
  \label{eq:h_i_regularized}
  h_2(y) & = & \frac{1}{1+{\rm e}^{\eta}} \,
  \frac{1-{\rm e}^{y^2/\theta}}{1+{\rm e}^{y^2/\theta-\eta}}
  \,\,\,,\\
  h_3(y) & = & \frac{1}{1+{\rm e}^{\eta}} \,
  \frac{\left( 1+{\rm e}^{-\eta} \right) \,\left( 
  1-{\rm e}^{y^2/\theta} \right)\,+\,\left(1+{\rm e}^{y^2/\theta-\eta}
\right) y^2/\theta}{ \left( 1+{\rm e}^{-\eta} \right) \,\left( 
  1+{\rm e}^{y^2/\theta-\eta} \right)}
  \,\,\,,\\
  h_4(y) & = & \frac{{\rm e}^{\eta}}{2 \left( 1+{\rm e}^{\eta}
    \right)^3 \left({\rm e}^{\eta}+{\rm e}^{y^2/\theta} \right)}
  \left[ 2\,-\,2 {\rm e}^{2 \eta+y^2/\theta}\,+\, {\rm e}^{\eta}
  \left( 4+ 2 y^2/\theta-y^4/\theta^2 \right)
  \right. \nonumber \\  & & \left. 
    -{\rm e}^{y^2/\theta} \left( 2-2 y^2/\theta +y^4/\theta^2
    \right)\,+\, {\rm e}^{\eta+y^2/\theta} \left(-4 +2
      y^2/\theta+y^4/\theta^2 \right) \,+\,
   {\rm e}^{2 \eta} \left( 2+ y^2/\theta +y^4/\theta^2\right)
   \right] 
  \,\,\,,\\ 
   h_5(y) & = & \frac{ {\rm e}^{\eta} }{ 6 \left(1+{\rm e}^{\eta} \right)^4
     \left( {\rm e}^{\eta}\,+\, {\rm e}^{y^2/\theta} \right)}\,
     \left[ -6 \left( 1+ {\rm e}^{\eta} \right)^3 \left( -1 + 
        {\rm e}^{y^2/\theta}  \right) \right. \nonumber \\
     & & \left. +6 \left( 1+ {\rm e}^{\eta} \right)^2 
     \left( {\rm e}^{\eta}+{\rm e}^{y^2/\theta} \right) y^2/\theta
     \,+\, 3 \left( -1+{\rm e}^{2 \eta} \right) 
     \left( {\rm e}^{\eta}+{\rm e}^{y^2/\theta} \right) y^4/\theta^2\,+\,
     \left( 1-4 {\rm e}^{\eta}+{\rm e}^{2 \eta} \right) y^6/\theta^3 \right] 
    \,\,\,,
\end{eqnarray}
with $H_i=\int_0^{\infty}\!dy\, h_i(y)/y^{2 i-2}$.
Also, for evaluating the integrals involved, it is convenient to
determine the expansion of each integrand for small $x$ by reading
off the corresponding terms in
Eq.~(\ref{eq:Fermi_function_zero}). Note, 
that the limits of $H_i$ in the non-degenerate and highly degenerate case are
analytically known from the corresponding expansions of 
Eq.~(\ref{eq:gx_non_degenerate}) and Eq.~(\ref{eq:gx_degenerate}), respectively.
Note also, that the values of $H_1$ and $H_2$ are given by Arista and
Brandt \cite{Arista84}, while all higher expressions are given here
for the first time.

\section{Large $x$ expansion coefficients}
\label{sec:appendix_large_x}

The Fermi integrals of order $l/2$ involved in the large $x$ expansion have been
determined by MATHEMATICA \cite{Mathematica6} exploiting the relation
of the Fermi integral to the polylogarithmic function \cite{Abramowitz72}
\begin{eqnarray}
  \label{eq:Fermi_Polylog}
  F_{\alpha}(x) & = & -\Gamma(1+\alpha) \, \mbox{Li}_{1+\alpha}\left( 
  - {\rm e}^x
  \right)\,\,\,.
\end{eqnarray}
The Fermi integrals $F_{-1/2},F_{1/2},F_{3/2},F_{5/2}$ 
can be easily obtained with the
interpolation formulas given by Antia \cite{Antia93}.

\section{Modified Q-D algorithm}
\label{sec:appendix_qd_algorithm}

We summarize the modified Q-D algorithm introduced by 
de Andrade et al. \cite{Andrade03}. We start with the series
expansions Eq.~(\ref{eq:expansion_at_zero}) and 
Eq.~(\ref{eq:expansion_at_infinity}) and allow for zero coefficients $\mu_i$.
Next, an auxiliary expansion is introduced by subtracting the expansion of the 
function $K/(1+z)$ with a free parameter $K$ from the above given
expansion generating a new set of coefficients $\gamma_i$. 
For a convenient choice of $K$, all $\gamma_i$ will be non-zero and 
the ordinary Q-D algorithm can be applied. 
The continued fraction representation 
\begin{eqnarray}
  \label{eq:continued_fraction_alpha_beta}
 & & \polter{\mu_0}{z-\beta_1^{(0)} }\,-\, 
     \polter{\alpha_2^{(0)} z}{z-\beta_1^{(0)}}\,-\, 
     \polter{\alpha_3^{(0)} z}{z-\beta_3^{(0)}} \,-\, 
     \polter{\alpha_4^{(0)} z}{z-\beta_4^{(0)}}\,-\, \dots\,\,\,,  
\end{eqnarray}
having the original series expansions Eq.~(\ref{eq:expansion_at_zero}) and 
Eq.~(\ref{eq:expansion_at_infinity}) can be reconstructed from the
coefficients $\gamma_i$ by the following algorithm:

\begin{itemize}

\item Determine $\gamma_i$ from a given set of $\mu_i$ and a convenient 
      choice of $K$ as
      \begin{eqnarray}
        \label{eq:calculate_gamma}
        \gamma_i & = & \mu_i\,+\,\left(-1\right)^{i+1} K\,\neq\,0 \,\,\,.
      \end{eqnarray}

\item Perform the Q - D algorithm
  \begin{eqnarray}
    \lambda_i^{(r)} \,=\,
    \delta_{i-1}^{(r+1)}\,+\,\lambda_{i-1}^{(r+1)} -\delta_{i-1}^{(r)} 
    & , & \delta_{i}^{(r)}\,=\,\lambda_{i}^{(r)} \delta_{i-1}^{(r-1)} /
                              \lambda_{i}^{(r-1)}\,\,\,, 
  \end{eqnarray}
  with the initialization
  \begin{eqnarray}
    \label{eq:qd_initialize}
    \lambda_1^{(r)} \,=\, 0 & , & \delta_1^{(r)} \,=\,\frac{\gamma_r}{
    \gamma_{r-1}} \,\,\,, 
  \end{eqnarray}
  and $i=2,3,\dots$, $r= \dots,-2,-1,0,1,2,\dots$.

\item Determine for $i=1,2, \dots$  
   the following quantities leading to $\alpha_i$ and $\beta_i$,
   \begin{eqnarray}
     \label{eq:modified_qd}
     b_{i,i-1} & = & b_{i-1,i-2} \,-\,
                    \left( \alpha_i+\beta_i \right) \nonumber \,\,\,, \\
     d_{i,i-1} & = & d_{i-1,i-2} \,-\,
                    \left( \lambda_i^{(0)} + \delta_i^{(0)} \right)
                   \,\,\,, \nonumber \\
     u_{i+1} & = & u_i \lambda_{i+1}^{(0)} \,\,\,,\nonumber \\
     d_{i+1,0} & = & -d_{i,0} \delta_{i+1}^{(0)} \,\,\,,\nonumber \\
     w_{i+1} & = & u_{i+1}\,+\,w_i \left( 
                  b_{i,i-1}-d_{i-1,i-2}-1 \right)\,-\,
                  u_i \left( d_{i,i-1}-b_{i-1,i-2}-1 \right) \,\,\,,\nonumber \\
     b_{i+1,0} & = & d_{i,0} \left[ 1 - \frac{u_{i+1}}{w_{i+1}} 
                            \left( 1\,-\, \frac{ b_{i,0}}{ d_{i+1,0}} 
                            \right) \right]^{-1}\,\,\,,\nonumber \\
     \alpha_{i+1} & = & \frac{w_{i+1}}{w_i} \,\,\,,\nonumber \\
     \beta_{i+1} & = & -\frac{b_{i+1,0}}{b_{i,0}}\,\,\,,\nonumber 
   \end{eqnarray}
   with the initialization
   \begin{eqnarray}
     \label{eq:alpha_beta_initialization}
     \alpha_1 \,=\,0 & , & \beta_1 \,=\, \frac{\delta_1^{(0)}}{
      \gamma_0-K \delta_1^{(0)}} \,\,\,, \nonumber \\
     w_1 \,=\,\mu_0 & , & u_1 \,=\,\gamma_0 \,\,\,, \nonumber \\
     b_{0,-1} \,=\, 0 & , & d_{0,-1}\,=\,0 \,\,\,.
   \end{eqnarray}
\end{itemize}
    
\section{Numerics of continued fractions}

There are several algorithms to numerically calculate the value of a 
continued fraction. Here, we use the backward recurrence algorithm
which is known for its numerical stability \cite{Lorentzen92}. 
Given $n$ partial numerators $a_i$ and partial denominators $b_i$, we
determine an approximation $t_0$ to the continued fraction by setting
$t_n$ to a small, non-zero number and perform the iteration
\begin{eqnarray}
  \label{eq:backward_algorithm}
  t_{i-1} & = & \frac{a_i}{b_i+t_i} \,\,\,,
\end{eqnarray}
 for $i=n, \dots, 1$.


\begin{thebibliography}{5}

\bibitem{Lindhard54} J. Lindhard, 
                     Mat.-Fys. Medd.-K. Dan. Vidensk. Selsk. \textbf{28}, (8)
                     (1954).

\bibitem{Mahan93}    G.D. Mahan, {\it Many-Particle Physics}
                     (Plenum Press, New York, 1993).

\bibitem{Ichimaru94}  S. Ichimaru, {\it Statistical Plasma Physics}
                      (Addison-Wesley, Reading, 1994).

\bibitem{Arista84} N.R. Arista and W. Brandt, Phys. Rev. A
  \textbf{29}, 1471 (1984).


\bibitem{Griem97}    H.R. Griem, {\it Principles of Plasma
                     Spectroscopy} (Cambridge University Press,
                   Cambridge, 1997).

\bibitem{Reinholz00}  H. Reinholz, R. Redmer, G. R{\"o}pke, and
                      A. Wierling, Phys. Rev. E \textbf{62}, 5648 (2000).




\bibitem{cont_phys} see e.g. H. Mori, Prog. Theor. Phys. \textbf{34},
                    399 (1965); M.H. Lee,
                    Phys. Rev. Lett. \textbf{49}, 1072 (1982);
                    J. Hor\'{a}\v{c}ek and T. Sasakawa, Phys. Rev. C
                    \textbf{32}, 70 (1985); D.E. Neuenschwander,
                    Am. J. Phys. \textbf{62}, 871 (1994). 


\bibitem{Baker96} G.A. Baker, {\it Pad\'{e} Approximants}
                  (Cambridge Univ. Press, Cambridge, 1996).    

\bibitem{Baker70} G.A. Baker and J.L. Gammel, {\it The Pad\'{e}
                  approximant in theoretical physics} 
                  (Academic Press, New York, 1970).

\bibitem{Thron48} W.J. Thron, Bull. Amer. Math. Soc. 
                  \textbf{54}, 206 (1948).


\bibitem{McCabe76} J.H. McCabe and J.A. Murphy, J. Inst. Maths Applics 
                   \textbf{17}, 233 (1976). 

\bibitem{Fried61} B.D. Fried and S.D. Conte, {\it The Plasma 
                      Dispersion Function} (Academic, New York, 1961).


\bibitem{Abramowitz72} M. Abramowitz and I.A. Stegun, 
                       {\it Handbook of Mathematical Functions}
                       (Dover, New York, 1972).

\bibitem{Henrici91} P. Henrici, {\it Applied and Computational Complex
                    Analysis, Vol. 2} (Wiley, New York, 1991)

\bibitem{Andrade03} E.X.L. de Andrade, J.H. McCabe, and A. 
                    Sri Ranga, J. Comp. Appl. Math. \textbf{156}, 487 (2003).

\bibitem{McCabe74} J.H. McCabe, Math. Comp. \textbf{28}, 811 (1974). 

\bibitem{McCabe84} J.H. McCabe, J. Plasma Physics \textbf{32},
                    479 (1984).


\bibitem{Mathematica6} Wolfram Research, Inc., Mathematica, Version
  6.0, Champaign, IL (2007).

\bibitem{Antia93} H.M. Antia, Ap. J Suppl. \textbf{84}, 101 (1993).

\bibitem{Lorentzen92} L. Lorentzen and H. Waadeland, {\it Continued
                      Fractions with Applications} (North-Holland,
                    Amsterdam, 1992).


\end{thebibliography}
\end{document}